\documentclass[twocolumn]{article}
\usepackage{graphicx}
\usepackage{amsmath}
\usepackage{authblk}
\usepackage{abstract}
\usepackage{footmisc}
\usepackage{cite}

\begin{document}

\title{Dynamically Sharded Ledgers on a Distributed Hash Table}

\author{Christoffer Fink}
\author{Olov Schelén}
\author{Ulf Bodin}
\affil{%
  Luleå University of Technology, Sweden
}


\twocolumn[
\maketitle
\begin{onecolabstract}

Distributed ledger technology such as blockchain is considered essential for supporting large numbers of micro-transactions in the Machine Economy, which is envisioned to involve billions of connected heterogeneous and decentralized cyber-physical systems. This stresses the need for performance and scalability of distributed ledger technologies. Addressing this, sharding techniques that divide the blockchain network into multiple committees is a common approach to improve scalability. However, with current sharding approaches, costly cross-shard verification is needed to prevent double-spending.
This paper proposes a novel and more scalable distributed ledger method named ScaleGraph that implements dynamic sharding by using routing and logical proximity concepts from distributed hash tables. ScaleGraph addresses cyber security in terms of integrity, availability, and trust, to support frequent micro-transactions between autonomous devices.
Benefits of ScaleGraph include a total storage space complexity of $O(t)$, where $t$ is the global number of transactions (assuming a constant replication degree). This space is sharded over $n$ nodes so that each node needs $O(t/n)$ storage, which provides a high level of concurrency and data localization as compared to other delegated consensus proposals.
ScaleGraph allows for a dynamic grouping of validators which are selected based on a distance metric.
We analyze the consensus requirements in such a dynamic setting and show that a synchronous consensus protocol allows shards to be smaller than an asynchronous one, and likely yields better performance.
Moreover, we provide an experimental analysis of security aspects regarding the required size of the consensus groups with ScaleGraph.
Our analysis shows that dynamic sharding based on proximity concepts brings attractive scalability properties in general, especially when the fraction of corrupt nodes is small.

\end{onecolabstract}

]

\footnotetext[1]{Corresponding author: \texttt{christoffer.fink@ltu.se}}

\section{Introduction}

The advancements in the Internet of Things (IoT), Cyber-Physical Systems (CPS), Artificial Intelligence (AI), and smart devices performing machine-to-machine (m2m) intercommunication drives the Machine Economy~\cite{khan2022review}. It is envisioned that Monetary compensations in the form of micro-transactions will involve billions of connected heterogeneous, decentralized, and diverse devices~\cite{saputhanthri2022survey}. Major pillars of the Machine Economy, besides IoT, CPS, and AI, include Distributed Ledger Technology (DLT)~\cite{johnk2021rise}. DLT such as blockchain serves as a decentralized backbone for supporting micro-transactions and payments at large scale. This stresses the need for scalability of DLT and blockchain to handle large numbers of micro-transactions and payments at low cost with minimal delay.

State-of-the-art approaches addressing DLT and blockchain scalability~\cite{8539529} can be categorized into first-layer solutions that propose modification within the main blockchain such as consensus and data storage, and second-layer solutions that propose modifications outside the chain~\cite{hafid2020scaling, monrat2023sharding}. First layer solutions include sharding~\cite{liu2023survey, baniata2023distributed, qi2023stfm, chen2023scaling}, bigger blocks~\cite{woznica2022performance}, alternatives to Proof-of-Work (PoW)~\cite{yadav2023comparative} and Directed Acyclic Graph (DAG) based solutions~\cite{zhang2023phantasm}. Second-layer solutions~\cite{gangwal2023layer2, gudgeon2020sok}, typically use a two-layer model with a second layer of parallel side chains (i.e., shard chains) that eventually commits or synchronizes with the main chain. There are also proposals for using two layers to accomplish a similar effect as first-layer sharding, where a coordinating main chain (sometimes known as a beacon chain) is connected to several shard chains (that include consensus committees). Then, transactions are dispatched to the shard chains for local consensus in parallel, with periodical synchronization back to the beacon chain to merge status across all shard chains~\cite{xu2023consensus, monrat2023sharding, Buterin2020Casper}.
Sharding, whether first-layer or second-layer solutions, means dividing the blockchain network into multiple committees, each processing a separate set of transactions.

Challenges of current sharding proposals include the need for cross-shard verification to prevent double-spending~\cite{8416236, Liu2020crossshard}. This is because each transaction involves at least two parties and consequently a transaction processed in one shard may have direct or indirect dependencies on transactions processed in other shards. Therefore, transaction verification in one shard may depend on transactions being verified or stored in another shard. This adds complexity to the solutions and limits the scalability. It is the combination of the need for transactions involving more than one party, the need for an immutable ordered history, and the need for consensus that together make sharding so hard in DLT\@. On the contrary, sharding in other kinds of distributed data stores can be effectively and efficiently provided. As an example, the Distributed Hash Table (DHT) provides completely dynamic and automatic sharding of data where each object can be immutable (and signed) per se, but then without any check of critical dependencies between objects, without maintaining an ordered history, and without any consensus provided. It may therefore seem that DLT addresses a completely different problem. Nevertheless, there are lessons to learn from the DHT\@.

The scope of this paper is to address the issue of cross-shard verification and to propose a novel DLT method named ScaleGraph with dynamic sharding using routing and logical proximity concepts from the DHT\@. ScaleGraph thereby combines the principles of DHTs with the principles of distributed ledgers (DLTs). Effectively, each transaction is processed for consensus, stored, and replicated in a subset of nodes that are logically close to the sender and the receiver respectively, and therefore the consensus committee has direct awareness of historical or concurrent transactions that could be conflicting. Therefore, cross-shard verification between the two shards concerned is done directly as part of the consensus decision.  This eliminates the need for hindsight cross-shard validation which is the norm in other approaches~\cite{chen2023scaling, liu2023survey, qi2023stfm, 8416236, Liu2020crossshard}. Global consistency is ensured by linking each transaction/block to both the sender and receiver history respectively as part of the consensus process, thereby forming a partially ordered DAG allowing maximum concurrency.

The main contributions of the paper are focused on first-layer sharding, including an analysis of requirements and a novel proposal for establishing sharding groups to obtain consensus. More specifically the contributions include:
\begin{itemize}

\item
An analysis of the critical event/transaction ordering to balance the trade-off between global synchronization and concurrent operation to provide both effective and efficient sharding (Section~\ref{sec:ordering}).

\item
A novel DHT-based sharding solution that maximizes parallelism by imposing a partial order (rather than total order) on transactions and validating them in nodes being logically close (as defined by the DHT distance metric) to the affected accounts (Section~\ref{sec:overview}). This includes defining a reliable consensus method for operation in such a dynamic setting of validators, based on state of the art.

\item 
A security analysis. First, a theoretical analysis of the implications of sharding, attacks, and consensus vote counting (Section~\ref{sec:security}).
Second, an experimental analysis of security aspects of the proposed solution w.r.t.\@ the required size of the consensus groups (Section~\ref{sec:experiment}).

\end{itemize}

The current version of ScaleGraph addresses cyber security in terms of integrity, availability, and trust, to serve as a secure critical infrastructure supporting autonomous systems that perform frequent micro-transactions between them. This includes an efficient, resilient, and trustworthy consensus model. Our evaluation shows the needed number of validator nodes in each shard vs the number of traitors. The security and trust numbers can be interpreted both in unpermissioned and permissioned scenarios. However, if high performance and resource efficiency (environmental sustainability) are prime objectives, there are benefits from having some permission control. By vetting the validator nodes and their governance, traitors can be prevented and tracked, thereby lowering the required degree of replication and size of consensus groups. Vetting and permission of validator nodes is however out of scope in this paper.

Like in any typical blockchain cases, the information that is validated and stamped by consensus must be available to validators. However, parts of the payload can be confidential and private information, typically encrypted (traditionally or homomorphically) and signed for immutability. It can be easily checked that the information is intact, and if desirable, who signed such information (but that can equally well be confidential). Since Scalegraph performs data sharding, it should be feasible to store bulky confidential information on the ledger, not requiring it to be stored off-chain. This is out of scope in this paper.

Benefits of ScaleGraph include a total storage space complexity of $O(t)$, where $t$ is the global number of transactions (assuming a constant replication degree). This space is sharded over $n$ nodes so that each node needs $O(t/n)$ storage.
The consensus decision and storage are delegated to $2r$ validator nodes where $r$ is a system-wide target replication number. The set of validators is different for combinations of sender and receiver address but may be overlapping and will change with node churn. Therefore, non-conflicting consensus decisions are run concurrently, typically by different validator nodes. This provides a high level of concurrency and data localization as compared to other delegated consensus proposals.

\section{Critical event ordering}\label{sec:ordering}

A key objective of consensus is to maintain a global order of events that are initially issued asynchronously.
A common model in blockchain is to enforce a total order of all events on an immutable and linear log that is replicated to a sufficiently large number of nodes to ensure that the global order can resist a certain number of Byzantine nodes.
However, global total ordering and horizontal scaling are inherently conflicting goals, because the latter objective is to have nodes operating independently of each other in parallel.
Therefore, to provide scale-out properties such as data sharding and independent concurrent consensus, the ordering constrains must be relaxed from ensuring a total order of all events to some partial order.
The less strict the imposed ordering constraints, the better the opportunity to shard and parallelize.
This is a fundamental trade-off.
The required constraints for event ordering are highly application-dependent and, therefore, there is no single best solution.
If the trade-off is well understood, a sufficient ordering constraint can be imposed to meet the application requirements, while still allowing for scalability and performance.

Now, which are the minimal ordering constraints?
In the simplest form of crypto currencies, it is strictly necessary to order spending events from each account in order to resolve double spending.
It is also strictly required to, for each account, order the reception events in relation to the spending events, because it is a requirement that spending events refer to some previously received currency transactions or to a correct balance.
Preventing double spending means that received transactions (or available balance) can be spent only once.
In case of conflicting transactions, where multiple spending events refer to the same previously received event (or available balance), at most one of the transactions can be valid.
The consensus process is responsible for deciding which spending event is eligible, and reject the others.
By contrast, reception events are not dependent on each other, assuming incoming transactions are always acceptable (or reception constraints do not depend on history).
Hence it is not strictly necessary to order them relative to each other, only relative to spending events.

From this we conclude that minimal ordering constraints in typical crypto currencies require per-account total order of spending events, and that the reception events must be ordered in relation to the spending events.

\begin{figure}
\centering
\includegraphics[width=\columnwidth]{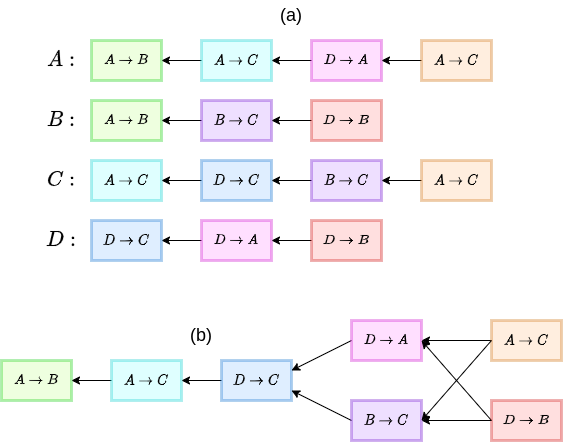}
  \caption{(a) shows a collection of block chains, each associated with one of the accounts A, B, C, and D\@. (b) shows the DAG representing the partial order of all the transactions in those chains. Since transactions/blocks appear twice in (a), once in each chain for the two affected accounts, they have been color coded to make the pairs easier to see.}\label{fig:blockdag}
\end{figure}

Note that, from the imposed per-account total order of events, it is at any time possible to extract the complete partial order of event histories involving any set of accounts.
That is, it is possible to combine the cross-account events into a directed acyclic graph (DAG) for any set of accounts.
This is illustrated in figure~\ref{fig:blockdag}, which shows a collection of individual block chains and the corresponding DAG\@.

\section{Overview of ScaleGraph}\label{sec:overview}

ScaleGraph is a network of nodes that implements a fully sharded blockchain system.
That is, both data storage (the ledger) and transaction validation are sharded, and data are replicated within each shard.
In particular, the distributed hash table (DHT) Kademlia~\cite{Maymounkov:2002:KPI:646334.687801} is used for defining shards, i.e.\@ the subset of nodes that belong to each shard.
For each account and associated blockchain, there is one shard consisting of $r$ nodes.
Client accounts and node IDs share the same address space.
The $r$ nodes that have IDs closest to an account form one shard and are responsible for storing and validating the transactions that involve that account.
Kademlia allows this set of nodes to be found in $O(\log N)$ steps.

A transaction has a sender and a receiver account.
To validate a transaction, the $r$ nodes closest to the sender account and the $r$ nodes closest to the receiver account, i.e.\@ the two shards, together form a validator group.
This validator group runs a Byzantine fault tolerant (BFT) consensus protocol to agree on the next block to add to both the sender and receiver chains.

Relaxing the ordering constraint to imposing a partial order rather than a total order on transactions makes transactions between different accounts independent of each other.
This allows any number of transactions to be validated in parallel as long as they involve different accounts.
Transactions are also processed individually rather than in blocks, which means there is no need for cross-shard validation.
More precisely, there are always exactly two shards involved in validating a transaction.

\begin{figure*}
\centering
\includegraphics[width=\textwidth]{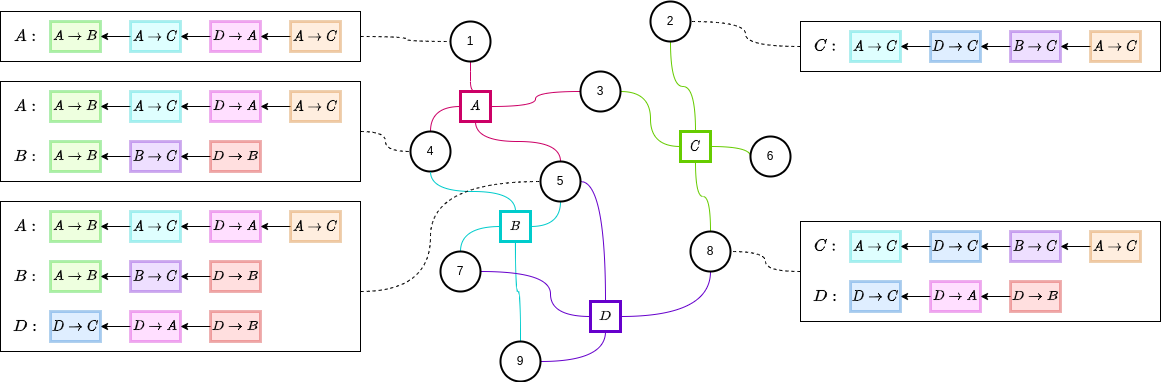}
  \caption{The squares represent accounts A, B, C, and D. The circles represent validator nodes. Lines between accounts and nodes show which nodes form shards associated with the respective accounts. The large rectangles show the block chains stored at some of the nodes. Since the same transaction/block appear multiple times, they have been color coded to make the duplicates easier to identify. See also Figure~\ref{fig:blockdag}, which more cleanly shows the block chain for each account. Each account is connected to 4 nodes, implying a shard size of 4.}\label{fig:scalegraph}
\end{figure*}

Figure~\ref{fig:scalegraph} shows an illustration of a ScaleGraph network.
Every node connected to an account stores a copy of the block chain for that account.
We can see that the shards for accounts $A$ and $B$ overlap, since they both contain nodes $4$ and $5$.
One of them, node $5$, is also a member of the shard for account $D$.
Hence that node must store the block chains for all three accounts.
Note that this illustration uses Euclidean distance in the plane.
The actual Kademlia distance metric would be more difficult to visualize.

\subsection{Kademlia as a DHT}

In Kademlia, nodes and data (key-value pairs) are assigned an address (an ID or a key) from the same $B$-bit address space.
Each node maintains a routing table of size $O(\log N)$, where $N$ is the number of nodes in the network, and $k$ is the replication factor, a global system parameter.
When Kademlia is used as a DHT, nodes want to do one of two things.
Either they want to store data in the network, or they want to find data in the network.
In both cases, this involves finding the $k$ nodes that have IDs closest to the key, which is usually the hash of the data.
The distance metric used to define logical distance between addresses is the \texttt{xor} function, with the result interpreted as a $B$-bit integer.
Where the bits of the two addresses match, the \texttt{xor} will have a $0$-bit.
A long common prefix therefore implies an \texttt{xor} with many leading zeros, and therefore a small distance.

The same basic lookup procedure is used both to find nodes and data.
When storing data, a node must find the $k$ nodes closest to the key, so that \texttt{store} requests can be sent to them.
To find this set of nodes, the querying node builds and iteratively refines a pool of candidate nodes, which will eventually contain the $k$ closest nodes.
The node initializes the pool by finding, in its local routing table, the $k$ nodes that are closest to the key.
It then sends \texttt{find-node} requests to a small number of those nodes in parallel.
When a node receives a \texttt{find-node} request, it replies with the $k$ closest nodes it can find in its routing table.
This allows the querying node to receive better (closer) candidates, which it can then query to find still better candidates, and so on, until it is no longer able to find new nodes that are closer to the key.
At this point the node issuing the request has found the $k$ nodes closest to the key and can ask them to store the value.

Retrieving data starts exactly the same way, except the node sends \texttt{find-value} requests instead of \texttt{find-node} requests.
When a node receives a \texttt{find-value} request, it either has a matching key-value pair stored, or it does not.
If it does, then it simply replies with the requested data.
Otherwise, it replies with the $k$ closest nodes it knows about, just as if it had received a \texttt{find-node} request.
The requesting node again receives better and better candidates until it either finds the value or is forced to conclude that the value does not exist in the network.

Kademlia normally stores immutable key-value pairs that expire after they have not been requested or refreshed for a certain amount of time, effectively implementing a kind of garbage collection.
This means that requests keep data alive.
In the absence of naturally occurring requests, the node that is asking the network to store a value must periodically refresh it to prevent it from expiring.
Nodes that store a key-value pair replicate it by periodically performing a \texttt{store} operation.
This ensures that close to $k$ copies always exist near the key so that redundancy is maintained even when nodes that used to store the data go offline or when new nodes appear near the key.
To avoid all $k$ nodes replicating data to the other $k-1$ nodes at the same time, nodes that receive a \texttt{store} should assume that its neighbors received one as well, and cancel its own replication timer.
The node that is the first to replicate is then the only one to do so.

\subsection{Adapting Kademlia to a Blockchain Application}

When using Kademlia in a blockchain application to store transactions, rather than as a DHT storing arbitrary key-value pairs, a few adaptations need to be made.

In the DHT case, Kademlia stores atomic values.
In the blockchain case, it would be straightforward to treat account addresses as the key, and the corresponding transaction log as the value.
However, it would be better to allow nodes to \emph{request individual blocks}, or ranges of blocks, rather than treating the whole chain as one value and forcing nodes to retrieve all transactions.
This is particularly useful when a node has been offline for a limited amount of time, has missed a few blocks, and needs to catch up to its peers.
Nodes must also be able to request information about the last block in a chain.
So we need to augment Kademlia with a few additional request types.

When a node replicates a chain, it sends the last block to its peers.
Based on the block height, a node that receives the block can determine how many blocks it is missing and request them to fill the gap.

The \emph{expiration time} should be much longer, so that transactions are only dropped by a node when it is clear that they no longer need to be stored by that node.
Assuming low churn (the rate at which nodes join and leave the network), the storage overhead resulting from a long expiration time would not be an issue.
With high churn and a long expiration time, a node may be a member of many shards and thus accumulate many chains before they expire.

In addition, \emph{expiration must be treated slightly differently} compared to the DHT case.
Unlike a key-value pair that has not been accessed for some time, transactions should never be deleted from the network, even when an account is inactive for an extended period of time.
So the standard expiration timer should be infinite.
Instead, what we care about is a node that no longer needs to store the chain for a particular account because it is now too distant from the account.
Every node can find out what the $k$ closest nodes to an account are, and, in particular, whether it itself is one of them.
When a node is about to replicate a chain that it is storing, it should check whether it is one of the $k$ closest nodes.
If it is, then it replicates the chain as usual.
If it is not, it replicates the chain a limited number of times.
After the final replication, it starts a timer and eventually drops the chain.

The appropriate \emph{replication period} would depend on the expected amount of churn.
For a stable network with little churn, replication could happen infrequently, similar to the DHT case.
With higher churn, a shorter period would be advisable.
An obvious reason is that it is important that transactions are not inadvertently lost.
A less obvious reason is that a long replication period would increase transaction latency.
Before a node can participate in consensus, it must have a copy of the transactions for the affected account.
If a node has recently become part of the validator group, either because it recently joined the network or because it effectively moved closer to the account address when another node left, it would first have to retrieve the transactions from its neighbors before it can participate in consensus.

\subsection{Sync HotStuff Consensus Protocol}

Once a transaction has reached a validator group, the nodes use a BFT consensus protocol to validate the transaction.
This is explained in detail in the next section.
By using a variant of Sync HotStuff~\cite{abraham2020sync}, a synchronous consensus protocol, each shard can tolerate up to half of a shard consisting of Byzantine nodes.
An asynchronous (or partially asynchronous) protocol can only tolerate up to a third.
As sections~\ref{sec:security} and~\ref{sec:related} elaborate on further, and as will be evident in the experimental results~\ref{sec:experiment}, this choice makes a crucial difference when determining the appropriate shard size.
A synchronous protocol allows shards to be much smaller, without sacrificing the fault tolerance of the overall system.

\section{Consensus in ScaleGraph}\label{sec:consensus}

When validating a transaction, Kademlia is used to find the nodes that should perform the validation.
The validator group consists of nodes close to the sender and receiver accounts affected by the transaction.
Once the transaction has reached the relevant nodes, they use a consensus protocol to agree on the next block to add to the sender and receiver account block chains.
We adopt Sync HotStuff as the consensus protocol but have to make some minor modifications.
The set of nodes participating in consensus depends on the transaction being processed and is therefore not known a priori~\footnotemark{}.
Therefore, proposals must include the list of validator nodes that are to participate.
For the purpose of counting votes, nodes are not generic, but must instead be treated as belonging to two different groups.

\subsection{The consensus process in detail}

\begin{itemize}
	\item
		Each validator node has a cryptographic key pair.
		The node is identified by its public key, and the private key is used for signing all messages, in particular proposals and votes.
	\item
		There is a configurable global parameter $r$.
    The $r$ nodes closest to a given key/ID (based on the Kademlia distance metric) are referred to as an $r$-group.
    The $r$-group on the sender side (i.e.\@ the $r$ nodes closest to the sender address) and the receiver side are referred to as the $r_S$-group and $r_R$-group, respectively.
  \item
    Together, the union of the two (possibly overlapping) $r$-groups forms a validator group, $V$, with $r \leq |V| \leq 2r$.
  \item
    Each $r$-group has a leader.
    The leader on the sender side, $L_S$, is responsible for making proposals during the consensus process.
    The leader on the receiving side, $L_R$, plays a less important role.
	\item
		Transactions contain the sender address (the public key of the account from which money is transferred) and the receiver address (the public key of the account to which the money is transferred) and are signed with the private key of the sender.
  \item
    Using the sender and receiver addresses in a transaction, any node can determine the members and ordering of the validator group based on the distance metric.
  \item
		Clients send transactions to any validator node.
		The node that receives the transaction forwards it to the appropriate validator group $V$, i.e.\@ the sender and receiver $r$-groups, $r_S$ and $r_R$.
  \item
    The first node in $r_S$ is expected to become the leader (a.k.a.\@ proposer) in the consensus process.
  \item
    There is a single transaction per block.
    A block consists of a transaction, a list of public keys of the nodes $V$ that should participate in consensus, and the hash and height of the last block on both the sender and receiver side.
    For increased accountability, a block may also include the list of validator votes for the previous block.
  \item
    Finding the height and hash of the last block of the chain on the sender side is easy, since $L_S$ already stores the entire chain.
    For the sender side to acquire the height and hash of the last block, $L_S$ sends a request to $L_R$, including the sender account address in the request.
    When $L_R$ replies, it also locks on the sender account until the transaction is either committed or expires.
  \item
    The leader puts the transaction in a block, together with the list of validators and the hashes and heights of the last blocks.
    The block is ultimately signed and sent to the other validator nodes as a proposal.
  \item
    At this point, our modified Sync HotStuff protocol takes over.
\end{itemize}

\subsection{Modified Sync HotStuff}

A node is leader/proposer for a period called a \emph{view}.
Views are numbered incrementally.
When a leader fails, a new leader must be appointed, also known as a \emph{view change}.
As in Sync HotStuff, a \emph{quorum certificate} $C_v(B_k)$ is a set of signed votes for a block $B_k$ from a quorum of nodes in view $v$.
We require at least $q = \lfloor \frac{1}{2}r \rfloor + 1$ votes from nodes in $r_S$ and $r_R$, respectively.
The reason for this more specific requirement on votes is explained in detail in Section~\ref{sec:votes}.
In a nutshell, we need to reach consensus in $r_S$ and $r_R$ in parallel since they each have access to different information and can only validate one side of the transaction.

In the following step-by-step description, the current view number, $v$, is included in every message.
Every message is also signed by the sender.
So view numbers and signatures are not mentioned explicitly.
Note that the timer in step 3 is non-blocking.

\begin{enumerate}
  \item
    After receiving a quorum certificate $C_v(B_{k-1})$, i.e.\@ a confirmation that the previous block $B_{k-1}$ was committed, $L_S$ constructs the next block, $B_k$.
    $L_S$ then builds a proposal containing the block $B_k$ and the certificate $C_v(B_{k-1})$ and broadcasts it to $V$.
  \item
    When a node receives a valid proposal, it forwards it to the other nodes and broadcasts a vote for $B_k$.
  \item
    Upon receiving $q$ proposals for block $B_{k+1}$ from each $r$-group, it starts a $\text{pre-commit-timer}_{v,k} = 2\Delta$.
    When the timer is triggered, the node pre-commits $B_k$ and broadcasts a commit message for $B_k$.
  \item
    When receiving $q$ commit messages for $B_k$ from each $r$-group, the node commits $B_k$ and all its ancestors.
\end{enumerate}

\subsection{Validator Groups and Leaders}

When the transaction from the client is initially received by a node, it must be forwarded to the whole validator group, rather than just $L_S$, so that the other validators know to expect a proposal, and a leader/view change can take place in case $L_S$ does not make the expected proposal in time.

While it is accurate to say that a node is a leader for a validator group, in the sense that it makes proposals in a consensus protocol that is executed by that group, the more important factor is that it is the leader for a particular sender address.
This is an important distinction, since there is considerable overlap between different validator groups.

A typical node will be involved as a validator in multiple validator groups.
In particular, it may be a leader in one group and a non-leader in one or more other groups at the same time.
Because each transaction involves only one sender, the same node can participate in multiple instances of the consensus processes simultaneously.
\emph{Any number of transactions between disjoint combinations of sender and receiver accounts can be processed in parallel.}

In fact, a node may even be the leader in multiple different validator groups simultaneously.
This would be the case if the node happens to be the closest node to two or more accounts, which means it could be a relatively common occurrence.
It could also happen if $L_S$ is replaced as leader, for example because it crashed, and the next node in line is $L_S'$, which is already the leader in another $r$-group (associated with a different sender account) and may even be engaged in a consensus process.
Either way, this scenario is benign, since there is nothing stopping the node from acting as leader for both senders.
The two roles involve two different sender accounts and therefore two different chains, allowing the node to lock on two different blocks.

A more unusual corner case to consider is when two different nodes are (briefly) the leaders for the same sender address.
This could happen if $L_S$ has just initiated the consensus process and a new node $L_S'$ that is closer to the sender address joins.
Before consensus is reached, the new leader could start processing another transaction.
This proposal would be ignored by the other validators while the current consensus process is still ongoing.

\subsection{Preventing or Resolving Deadlock}

When concurrent transactions do not have disjoint combinations of sender and receiver, there is a risk of wasting resources on retrying transactions under conditions of high contention.
Suppose many transactions involve the same receiver account.
If $L_R$ would not pace its response and lock on the next block, awaiting a matching proposal, but instead reply to any request for information on the last block, then the information it provides would only be valid for one of the pending transactions.
Once one transaction completes, the last block changes.
The other transactions would fail and have to restart, requiring the last block to be requested anew.
This retrying would waste resources and introduce unnecessary delays, and so locking improves efficiency.
However, it also introduces the possibility of deadlock.

Suppose there are three transactions involving accounts $A$, $B$, and $C$.
In particular, say we have transactions $A \to B$, $B \to C$, and $C \to A$, forming a cycle.
This means that there are three nodes that act as both $L_S$ and $L_R$ simultaneously.
Now suppose each of the three sending-side leaders simultaneously initiate the consensus process by requesting information from the receiver side.
In their role as $L_R$, they are already busy with one transaction and have to delay their reply until the pending transaction is concluded.
Hence they are deadlocked.

One possible approach is to assume that such an event will be exceedingly rare.
The typical case could then be handled optimistically.
Should deadlock occur, it can be resolved with a timeout, forcing the transactions to be retried, possibly with a randomized back-off timer to prevent all transactions from restarting simultaneously.
This would be a reasonable solution in a permissioned environment.

A more proactive solution would be to take advantage of the fact that accounts have a total order, and lock on either the sender or the receiver side first, depending on which of the two addresses is smaller.
Either $L_S$ locks on the next block \emph{before sending} a request to $L_R$, or it locks after \emph{receiving the reply}.
An arbitrary convention is to lock before the request if the sender address is smaller, or after receiving the reply if the receiver address is smaller.

A potential downside of the proactive approach is that accounts with a smaller ID receive preferential treatment.
Accounts with large IDs may experience starvation in extreme circumstances.
A longer cycle, involving a larger number of nodes and transactions, could introduce sufficiently long delays that one or more of the last transactions in line expire before they can be processed, and they would then have to be retried, adding further delays.
However, as the cycle length and thus delay increases, their probability also decrease.
A perfectly timed batch of transactions forming a long cycle is highly improbable.

\section{Security Analysis}\label{sec:security}

By implementing an \emph{immutable} ledger, blockchains excel at integrity.
They make it next-to impossible to rewrite history.
So two remaining considerations are, first, how to ensure that only valid information is added to the ledger in the first place (e.g.\@ preventing double spending) and, second, that the data remain available.
These questions ultimately come down to the probability that an adversary can control more nodes in a shard than what the consensus protocol can tolerate.

\subsection{Security implications of sharding}

Sharding introduces a trade-off between performance and security.
The nodes are divided into smaller groups that can process transactions in parallel, increasing throughput.
But this means that each group consists of a much smaller number of nodes, $r$, compared to the total number of nodes in the network, $N$.
Hence an attacker only needs to control a fraction of $r$ nodes rather than a fraction of $N$, which means that a smaller number of compromised nodes is sufficient for an attack.
Even assuming that an adversary can only gain control over randomly distributed nodes, sharding inherently lowers the total resilience of the network.
In other words, the attacker only needs to control a majority of some small, local subset of nodes, rather than a global majority.

As a concrete example, suppose there are $N = 15$ nodes and the consensus protocol can tolerate $b = 7$ Byzantine nodes, i.e.\@ $f < 1/2$.
Now suppose the nodes are divided into $m = 5$ shards, each of size $r = 3$.
Since $7/5 > 1$, there is at least one shard that has at least $2$ Byzantine nodes, and so that shard is compromised.
In general, because each of the $m$ shards must contain at least $\frac{r}{2} + 1$ honest nodes, i.e.\@ a majority, that means the whole network must contain at least $\frac{N}{2} + m$ honest nodes.
So, with sharding, it is no longer sufficient to have just $\frac{N}{2} + 1$ honest nodes.
Now suppose $b = 3$; a much smaller fraction of the nodes are Byzantine.
It is no longer guaranteed that a shard must necessarily be compromised.
However, each shard can be viewed as a random sample.
Hence each shard represents an opportunity to achieve a local majority.
As shards are made smaller and more numerous, the probability that at least one shard is compromised increases.
A quantitative analysis of this probability, and how it depends on parameters such as the shard size, can be found in~\cite{hafid2022tractable}.
It should be noted, however, that their results cannot be directly applied to ScaleGraph, since they are based on the assumption that shards are disjoint and thus sampling is done without replacement.
In ScaleGraph, shards are not disjoint.
In addition, shards do not consist of arbitrary subsets of nodes, but specifically of nodes that all have IDs close to some account address, and therefore to each other.

Because sharding divides nodes into smaller groups that are easier to take over, a consensus protocol that tolerates a fraction $f$ of Byzantine nodes within a validator group has a global resilience of $f' < f$ that is much smaller.
The global resilience can be increased by making the shards larger.
But dividing the network into larger (and therefore fewer) shards sacrifices performance by involving more nodes in the consensus protocol.
Using an underlying consensus protocol with a higher resilience, i.e.\@ a synchronous one, allows the global resilience to be higher without increasing the size of shards.
RapidChain~\cite{zamani2018rapidchain} used the same idea to achieve much higher resilience than Elastico~\cite{luu2016secure} and OmniLedger~\cite{8418625}, two earlier sharding blockchains.

However, ScaleGraph behaves slightly differently to typical approaches to sharding.
The difference is easiest to illustrate by considering what happens when there are few nodes relative to the number of accounts/shards.
In that case, nodes are massively recycled and included in multiple shards.
But this means that the number of shards has a much smaller influence on the global resilience.
Additionally, parallelism is still possible.
In fact, even if the number of nodes equals the shard size, such that every account has identical sets of validator nodes, the leader will typically vary.
This means transactions can still be processed concurrently.

Another aspect affected by sharding is availability.
Storing transactions for an account only in the associated shard, a subset of the nodes, obviously reduces the amount of redundancy compared to the unsharded case where every node stores every transaction.
This is an unavoidable trade-off, since reducing storage requirements via sharding and increasing redundancy are conflicting goals.
However, the trade-off is configurable, since the shard size can be set to strike the appropriate balance for a particular application.
In fact, it would be possible to define a separate replication parameter $k \geq r$ that can be used to increase storage redundancy without automatically enlarging the validator group. (Of course, $k < r$ would not make sense, since some validator nodes would not be storing the chains they are supposed to operate on.)

\subsection{Sybil attacks}

A Sybil attack~\cite{douceur2002sybil} involves creating multiple identities, allowing a single entity to pretend to be several, thus gaining undue influence.
For example, it would be easy to influence the outcome of an election if one could cast an unlimited number of votes, each from a fake identity.
In Bitcoin, this risk is mitigated by the proof-of-work mechanism, which means that there is a cost associated with running a fake identity, in the sense that there is no advantage to be gained from creating a fake identity if one does not also use it for hashing.
The limiting factor therefore becomes computational resources rather than the number of identities.
In a system that uses a classic BFT protocol involving voting, it is precisely the number of identities an adversary controls that matters.

If nodes could choose their own IDs, an attacker could strategically position itself around a particular account address and take over a validator group.
Even if attackers are limited to generating (or being assigned) uniformly distributed random IDs, the probability of being able to compromise \emph{some} validator group would increase with the number of IDs.
One possible solution is to impose a cost to generating an identity, for example via a proof of work.
However, over time, an adversary could keep slowly generating additional identities.

Sharding greatly exacerbates this problem because the validator groups are relatively small, certainly when compared to the unsharded case where there is a single large group.
All the attacker needs is for multiple random IDs to accidentally cluster somewhere in the address space.
Then a validator group for an account near that cluster can potentially be taken over.
This is one reason why it is common for sharding solutions to rebuild the shards (or committees), and requiring new identities to be generated for each epoch~\cite{luu2016secure, zamani2018rapidchain}.
This effectively resets the attacker's progress each epoch and limits the accumulation of identities.

In a permissioned setting, where IDs are assigned based on some kind of validation of a real-world identity, these problems are solved.
However, Byzantine fault tolerance is still necessary to guard against legitimate nodes being compromised by an attacker.

Another advantage of rebuilding shards periodically is that it prevents an adversary from targeting a particular shard in the long term.
But in ScaleGraph, shards are relatively static (except for churn) and defined by the nodes' IDs.
Since there is no procedure for building shards, there is no natural way to re-build them.
However, this problem could be solved by a standard certificate expiration mechanism so that nodes are required to renew their keys/IDs.
This would effectively move them around in the address space periodically and therefore change the configuration of shards.

\subsection{Double Spending and Vote Counting}\label{sec:votes}

A naive implementation of the consensus protocol for the validator group would severely degrade fault tolerance.
To prevent double spending attacks, special care must therefore be taken when counting votes.

The validator group (or committee) agrees on which block is to be the next block in the chain, which involves checking that the proposed block is valid.
Assuming shards do not overlap, the group consists of $|V| = 2r$ nodes.
However, the two $r$-groups have access to different information, and they each validate one side of the transaction.
On the receiver side, the nodes in $r_R$ agree that the block is valid, in the sense that, for example, the hash of the previous block is correct.
On the sender side, the nodes in $r_S$ must also validate the transaction in the context of previous transactions to prevent double spending.
Nodes in $r_R$ do not have access to the information that is necessary for performing this validation.
So, even if all nodes in $r_R$ are honest, they would always vote `yes' without examining the transaction, provided the block is otherwise valid.
In effect, honest nodes could vote for an invalid transaction.
This means that a relatively small number of corrupt nodes on the sender side could create a majority and get an invalid transaction committed.

For example, for a protocol such as PBFT, $\frac{2}{3} |V| + 1$ votes for a proposal are necessary to commit a block.
Since $r_R$ will contribute $|r_R| = \frac{1}{2} |V|$ votes, only $\frac{1}{3} |r_S| + 1$ additional votes are needed on the sender side.
Hence only a $\frac{1}{3}$ fraction of Byzantine nodes in $r_S$ can be tolerated, rather than $\frac{1}{3}$ in $V$.
As a fraction of the total number of nodes involved in consensus, that is $f < \frac{1}{6}$. 
The situation is worse for a protocol such as Sync HotStuff, which requires only $\frac{1}{2} |V| + 1$ votes to commit.
In that case, a single corrupt node on the sender side can cause an invalid transaction to be committed.

A solution is to explicitly treat the validator group as the union of the two $r$-groups and insist that there is a majority of votes from both $r$-groups.
That is, merely collecting $\frac{1}{2} |V| + 1$ votes in total may not be sufficient, and a more specific requirement is that there are at least $\lfloor\frac{1}{2} |r|\rfloor + 1$ votes per $r$-group, taking the group membership of each vote into account.
This also means that the presence of Byzantine nodes in the individual $r$-groups is what matters, rather than in the whole validator group.
(If at most half the nodes in each $r$-group are Byzantine, then it follows that at most half of the nodes in $V$ are Byzantine.)

\section{Simulation Experiment}\label{sec:experiment}

While it has been studied how sharding affects fault tolerance (see Section~\ref{sec:related}), it is not obvious to what extent those methods can be applied to ScaleGraph.
In particular, we perform a simulation experiment to investigate whether the hypergeometric distribution accurately predicts the probability that at least one shard is compromised.
We also find appropriate shard sizes for several scenarios and compare the effect the fault tolerance of the consensus protocol has on the overall fault tolerance of the network as a whole.

\subsection{Setup}

All experiments use a variation of the following general setup.
Each repetition of the experiment first creates a random network of nodes and accounts and then runs $5000$ iterations checking for compromised shards.
The network is created by generating $N$ uniformly distributed node IDs and then $m$ shards.
Shards are created by generating $m$ random account IDs and, for each one, finding the $r$ closest node IDs.
A parameter $F$ specifies the fraction of nodes that are Byzantine.
Each iteration randomly samples $b = FN$ out of the $N$ nodes to be Byzantine and then inspects each shard, counting the number of nodes that are in the Byzantine set.
If the number of Byzantine nodes in a shard exceeds the per-shard tolerance $f$ (e.g.\@ there are at least $r/2$), then the shard is compromised.
Some of the experiments run $20$ repetitions, for a total of 100,000 iterations, and some run $100$ repetitions, for a total of 500,000 iterations.

The simulation uses the following parameters:
\begin{itemize}
  \item $N$: nodes in the network
  \item $m$: shards/accounts in the network
  \item $r$: shard size
  \item $F$: fraction of Byzantine nodes in the network
  \item $f$: fault tolerance of the consensus protocol
\end{itemize}

\subsection{Finding Required Shard Sizes}

\begin{figure}
\centering
\includegraphics[width=\linewidth]{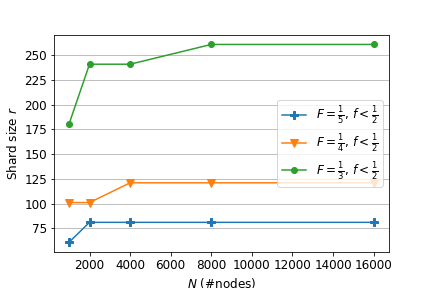}
  \caption{Required shard sizes for different network sizes and different fractions $F$ of Byzantine nodes present in the network, assuming a consensus protocol with $f < \frac{1}{2}$. Shard sizes for $F=\frac{1}{5}$, $F=\frac{1}{4}$, and $F=\frac{1}{3}$ are shown in blue (plus), orange (triangles), and green (discs), respectively. The number of shards is $m=2N$ and the results are based on $20 \cdot 5000 = 100000$ iterations.}\label{fig:r_for_N}
\end{figure}

Suppose there are $N$ nodes in total, out of which a fraction $F$ are Byzantine, and the consensus protocol can tolerate a fraction $f < \frac{1}{2}$ of Byzantine nodes within a shard.
This experiment is used to find the shard size $r$ that is sufficiently large such that there is not a single compromised shard in any of 100,000 total iterations.
The experiment is run as described above, with $m=2N$.
The number of nodes is varied from $1000$ up to $16000$, and the fraction of Byzantine nodes takes on the values $\frac{1}{3}$, $\frac{1}{4}$, and $\frac{1}{5}$.
Shard sizes are considered in steps of $20$.
If any shard in any iteration is compromised, the shard size is too small, and the procedure restarts with a larger shard size $r' = r+20$.
For $f < \frac{1}{2}$, the results are shown in figure~\ref{fig:r_for_N}.

To illustrate the impact of the fault tolerance of the consensus protocol, the experiment is also run under the assumption that only $\frac{1}{3}$ of Byzantine nodes can be tolerated (as would be the case with e.g.\@ PBFT) rather than $\frac{1}{2}$.
However, for $f < \frac{1}{3}$, some of the other parameter values must be excluded.
First, it is not possible to find a sufficiently large shard size for $F = \frac{1}{3}$, as there must necessarily be compromised shards.
(Note that this would be the case even for slightly smaller $F$, unless $r \approx N$.)
Second, because the shards are so large, the running time of the experiment becomes problematic for large $N$, and the larger values have therefore been omitted.
The range has instead been extended downward, starting at $N=125$ and going up to $N=2000$.

\begin{figure}
\centering
\includegraphics[width=\linewidth]{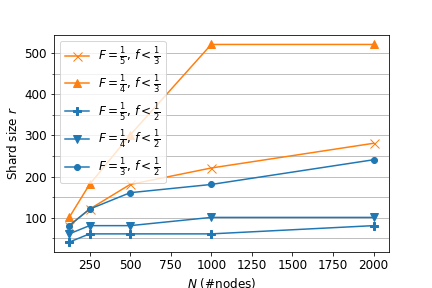}
  \caption{Comparing required shard sizes for different network sizes and different combinations of consensus protocol fault tolerances ($f < \frac{1}{2}$ in blue, $f < \frac{1}{3}$ in orange) and fractions of Byzantine node present in the network. The number of shards is $m=2N$ and the results are based on $20 \cdot 5000 = 100000$ iterations.}\label{fig:1_2_vs_1_3}
\end{figure}

We can see in figure~\ref{fig:1_2_vs_1_3} that the less resilient consensus protocol results in shards that must be more than twice as large, at least for small networks.
For example, with $N=250$ and $F=\frac{1}{4}$, we have $r=81$ (for $f<\frac{1}{2}$) versus $r=181$ (for $f<\frac{1}{3}$).
For slightly larger networks, the difference is even bigger.
For example, with $N=2000$ and $F=\frac{1}{4}$, we have $r=101$ (for $f<\frac{1}{2}$) versus $r=521$ (for $f<\frac{1}{3}$), more than a factor of $5$ difference.
Note that the shard sizes for $F=\frac{1}{3}$ and $f < \frac{1}{2}$ are smaller (or equal) compared to $F=\frac{1}{5}$ and $f < \frac{1}{3}$.
In other words, with the same (or smaller) shard size, we can achieve a global resilience of $F\leq\frac{1}{3}$ instead of $F\leq\frac{1}{5}$.

\subsection{Comparing Observed Probability to a Hypergeometric Distribution}

If shards were created by simply choosing random samples (with or without replacement) from a set of nodes that are equally likely to be chosen, then computing the probability of sampling at least one compromised shard would be straightforward.
Unlike sharding solutions with disjoint shards, ScaleGraph allows shards to overlap, which means nodes are replaced \emph{between} samples.
Of course, there is no replacement \emph{within} samples, since the same node is not a member of the same shard twice.
This suggests a hypergeometric distribution may potentially provide an adequate approximation.
However, that any node would be equally likely to be chosen for a shard is far from accurate, since shards consist specifically of nodes with IDs close to some account.
This experiment compares the observed probability of compromised shards, based on simulated sharding, to an estimate assuming a hypergeometric distribution.

To measure the probability, we run the experiment as described above.
The shard size is varied from $11$ to $101$ in steps of $10$, keeping all other parameters fixed.
The total number of nodes is $N=2000$, there are $m=2N$ accounts (and therefore shards), the number of Byzantine nodes is $b = \frac{N}{4}$, and the consensus protocol can tolerate a fraction $f < \frac{1}{2}$ of Byzantine nodes in a shard.
For each shard size, each repetition of the experiment counts the number of iterations in which at least one shard was compromised.
There are $100$ repetitions with $5000$ iterations for a total of $500,000$ iterations.
The probability is then the fraction of iterations with compromised shards out of all $500,000$ iterations.

Now for computing the estimated probability based on a hypergeometric distribution.
Let the random variable $X$ denote the number of Byzantine nodes that are included in a shard.
When there are $N$ nodes in total, out if which $b$ are Byzantine, and the shard size is $r$, then the probability of including exactly $k$ Byzantine nodes in a shard is
\begin{equation*}
  \Pr[X = k] = \frac{\binom{b}{k}\binom{N-b}{r-k}}{\binom{N}{r}}.
\end{equation*}
Since the consensus protocol can tolerate a fraction $f < \frac{1}{2}$ of Byzantine nodes, a shard is compromised for any $k = \lceil \frac{r}{2} \rceil, \lceil \frac{r}{2} \rceil + 1, \dots, r$.
So the probability of sampling a compromised shard is
\begin{equation*}
  p = \Pr\left[X \geq \frac{r}{2} \right] = \sum_{k=\lceil \frac{r}{2} \rceil}^r \frac{\binom{b}{k}\binom{N-b}{r-k}}{\binom{N}{r}}.
\end{equation*}
Hence the probability of sampling a shard that is \textbf{not} compromised is
\begin{equation*}
  1 - p,
\end{equation*}
and so the probability that none of $m$ shards are compromised is
\begin{equation*}
  {\left(1 - p\right)}^m.
\end{equation*}
Finally, the failure probability, i.e.\@ the probability that at least one of the $m$ shards is compromised, is therefore
\begin{align*}
  & 1 - {\left(1 - p\right)}^m \\
  ={\space}& 1 - {\left(1 - \sum_{k=\lceil \frac{r}{2} \rceil}^r \frac{\binom{b}{k}\binom{N-b}{r-k}}{\binom{N}{r}}\right)}^m.
\end{align*}

\begin{figure}
\centering
\includegraphics[width=\linewidth]{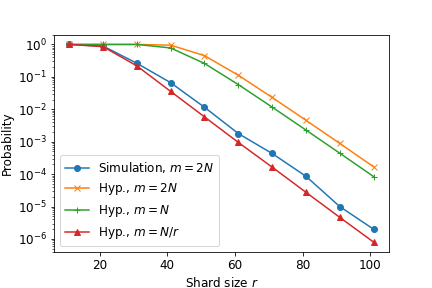}
  \caption{Failure probability (on a logarithmic scale) for different shard sizes. The parameters are $N=2000$, $F=\frac{1}{4}$, $f<\frac{1}{2}$, and $m=2N$.}\label{fig:p_for_r}
\end{figure}

Figure~\ref{fig:p_for_r} shows the observed failure probability, as well as the computed estimates, for different shard sizes.
The failure probability estimate has been computed for the same number of shards as in the simulation, but also for $m=N$ and $m=\frac{N}{r}$.
While $m = \frac{N}{r}$ is much smaller than the number of shards used in the simulation, the resulting probabilities agree reasonably well.
This adjusted shard count ignores overlap and matches the number of shards that would be present if they were disjoint.

\subsection{Effect of Number of Shards on Failure Probability}

\begin{figure}
\centering
\includegraphics[width=\linewidth]{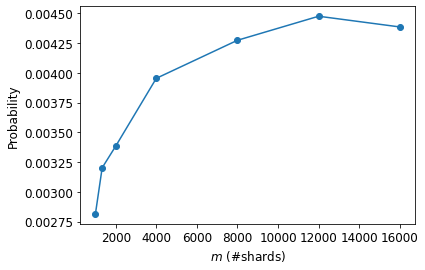}
  \caption{Failure probability for shard counts ranging from $m=1000$ to $m=16000$. Other parameters are fixed at $N=4000$, $r=61$, $F=\frac{N}{4}$, and $f < \frac{1}{2}$. Probabilities were found for each $m$ by simulating $100\cdot 5000=500,000$ iterations.}\label{fig:p_for_m}
\end{figure}

The number of shards directly depends on the number of accounts.
However, because there is typically extensive overlap between shards, it is not obvious how the number of shards/accounts affects the probability that at least one shard is compromised.
In this experiment, the number of shards is varied while keeping all other parameters fixed.
In particular, the number of nodes is $N=4000$, and $m$ takes on the values $\frac{N}{4}$, $\frac{N}{3}$, $\frac{N}{2}$, $N$, $2N$, $3N$, and $4N$.
The failure probability is computed as the fraction of iterations (out of all $100 \cdot 5000 = 500,000$) that have at least one compromised shard.
In order to get meaningful results, the shard size has been set to $r=61$ so that it is sufficiently small to allow compromised shards to occur.
Figure~\ref{fig:p_for_m} shows how the probability depends on $m$.
While there is clearly a tendency for a larger number of shards to result in a higher failure probability, the influence is fairly small.
Increasing shards by a factor of $16$ increases the probability by a factor less than $2$.

\section{Background and Related Work}\label{sec:related}

\subsection{Consensus Protocols Background}

Blockchain technology has sparked a renewed interest in Byzantine-fault tolerant (BFT) consensus protocols.
See~\cite{xu2023consensus} for a recent survey of consensus protocols for blockchain applications, or~\cite{brotsis2021suitability} for an IoT-centered perspective.
This section provides a background on consensus protocols and motivates our decision to use a synchronous protocol in general and a modification of Sync HotStuff in particular.

In a synchronous system, messages are delivered within a \textbf{known} time limit $\Delta$, and nodes in the network have synchronized clocks or communicate by rendez-vous.
This allows them to definitively detect crashed processes and to operate in lock-step.
However, an asynchronous system is a more realistic assumption, especially when considering real-world networks like the Internet, where messages can be arbitrarily delayed.
Unfortunately, the famous FLP-impossibility~\cite{fischer1985impossibility} result states that no deterministic protocol can reach consensus in an asynchronous distributed system if there is even a single node failure.
Some BFT consensus solutions, such as DLS~\cite{dwork1988consensus} and PBFT~\cite{castro99practical}, avoid the problem by assuming a \emph{partially asynchronous} system, as a middle ground between synchrony and asynchrony.
For example, one version of partial synchrony is to assume some finite but \textbf{unknown} bound on message delivery.
This allows the protocol to rely on synchrony only for liveness, which means progress can be made during periods where the network is functioning normally.
In a Byzantine setting, where nodes can behave maliciously, an attacker can delay progress indefinitely by carefully choosing when to respond to messages and when to remain silent~\cite{miller2016honey}.

Another strategy for circumventing the FLP-impossibility result is to introduce nondeterminism.
For example, HoneyBadgerBFT~\cite{miller2016honey} is a fully asynchronous BFT consensus protocol.
An advantage of an asynchronous protocol is that it need not rely on message timeouts and can make progress when messages are delivered.
Protocols that make progress as soon as the necessary messages arrive, without having to wait for a timeout, are called \emph{responsive}.
Typically, (partially) asynchronous consensus protocols are responsive.

However, there is one drawback with asynchronous BFT protocols, whether fully or partially asynchronous.
They can only achieve a resilience of $f < 1/3$, meaning they can only cope with a fraction of strictly less than a third of nodes being Byzantine.
In particular, adding authentication does not improve the resilience.
By contrast, synchronous protocols that use digital signatures can achieve a resilience of $f < 1/2$, i.e., consensus can be reached as long as less than half the nodes are Byzantine.
This potentially makes synchronous protocols attractive for use in sharding solutions.
There is a tension between wanting shards that are small for performance reasons, but large for security reasons.
If the consensus protocol has a higher fault tolerance, then shards can be made smaller, improving performance without sacrificing security, or, alternatively, improving security without sacrificing performance.

Historically, synchronous consensus algorithms have been slow.
For example, until a few years ago, the best synchronous protocol, due to Katz and Koo~\cite{katz2006expected}, required 24 rounds in expectation to reach consensus, and Abraham et al.~\cite{abraham2017efficient} improved this to an expected 8 rounds.
Another problem is that it is not clear what to do about a node that times out.
In other words, what happens when a node crashes and stops responding, or when an adversary changes the network parameters to violate the synchrony assumptions?
Guo~et~al.~\cite{guo2019synchronous} note that classical synchronous consensus protocols are underspecified and unimplementable in practice.
Recent work has addressed both problems.

Guo et al.~\cite{guo2019synchronous} introduce a relaxed synchrony assumption that takes into account that nodes may temporarily fail to respond on time.
This paper adopts the term \emph{mobile sluggish}, which was suggested by Abraham et al.~\cite{abraham2020sync} to describe this relaxed synchrony model to avoid confusion with preexisting models.
At any moment, some of the nodes may be \emph{sluggish}, which means they fail to respond on time.
Nodes that respond within the $\Delta$ bound are called \emph{prompt}.
The term \emph{mobile} is used to capture the fact that a different subset of nodes may be sluggish or prompt at any time, and the authors further assume that an adversary can choose which nodes will be sluggish.

In Sync HotStuff~\cite{abraham2020sync}, the performance problem is essentially eliminated by relying on synchrony only when committing a value (such as a transaction or block of transactions).
A timeout of $2\Delta$ is used for safety, ensuring that a conflicting proposal by the leader would have been received before committing.
But nodes do not block on this timeout and can start processing the next transaction while the timer is ticking, which means they can have multiple commit timers running simultaneously, creating a pipelining effect.
This means that nodes do not need to operate in lock-step, and \emph{the $\Delta$ time bound does not affect throughput in steady state}, no matter how conservative.

Furthermore, Sync HotStuff borrows an optimization from Thunderella~\cite{pass2018thunderella}: optimistic responsiveness.
As long as the actual fraction of Byzantine (or sluggish) nodes is $f < 1/4$, the protocol can run in a responsive mode, reducing the commit latency from $2 \Delta + O(\delta)$ to just $O(\delta)$, where $\Delta$ is the upper bound on message deliveries, and $\delta$ is the actual time to deliver a message.

While an asynchronous protocol could potentially achieve lower latencies, we have to remember that transaction processing also includes a routing or gossiping process in addition to the consensus protocol.
Kademlia performs node lookups in $O(\log N)$ steps, each requiring a round trip of $O(\delta)$.
So, for example, cutting the consensus latency in half would not halve the total latency.
In addition, the lower resilience of an asynchronous protocol implies larger shards, which also incurs a performance penalty.

\subsection{Sharding}

Full blockchain sharding has been proposed in RapidChain~\cite{Zamani:2018:RSB:3243734.3243853}, improving on earlier solutions like Elastico~\cite{luu2016secure} and OmniLedger~\cite{8418625}.
Several objectives are similar to ours.
They use a number of committees that operate concurrently for delegated consensus and inspiration from Kademlia is used for routing between committees for fast cross-shard verification.
However, major differences are as follows.
There is a reference committee that drives periodic reconfiguration of the consensus committees in epochs.
Besides changing the membership of the committees, also each member gets a fresh identity.
There is a peer-discovery mechanism where members of a committee can discover each other.
Each committee is responsible for maintaining a disjoint transaction ledger known as a shard.
Transactions are routed deterministically to the shards by hashing on the transaction ID\@.
Transactions are put in blocks.
Hashing will spread transactions over the shards so each time a committee shall check the validity of a transaction it has to reach out to other committees to verify that the referred-to input transactions exist in their shards.
To prohibit double spending, it should also be checked that none of those input transactions have already been consumed in any of the other shards.
Although the objectives are similar to ScaleGraph, this approach is very different, especially regarding forming and responsibilities of committees and the extensive communication between committees that is much less, or not needed at all between the validator groups of ScaleGraph.

In a scale-out blockchain for value transfer with spontaneous sharding~\cite{8525387}, it is shown that the complete set of transactions is not a necessity for the prevention of double-spending if the properties of value transfers is fully explored, and that a value-transfer ledger then has the potential to scale-out.
They provide sharding based on an off-chain structure, which contains individual chains for nodes to record their own transactions and a main chain for the consensus of the abstracts of their chains, i.e., provides a shared global state.
They assume a PKI infrastructure where nodes can link between identity and public key.
Each block contains multiple transactions and a hash digest of previous block.
Each node has an individual chain to record their own transactions in a FIFO fashion.
Periodically, so-called abstracts are sent to the main chain.
Nodes can acquire and cash chains from other nodes in order to validate transactions.
They show that if $N$ nodes are placed on a logical ring and each node receives transactions from the $c$ nearest node on the left and sends to the $c$ nearest nodes on the right.
The approach is very different from ScaleGraph and sharding relies on the said locality of connectivity, i.e., how many other nodes that each node communicates with.
If the number is too high, no sharding benefits are obtained.

PolyShard~\cite{DBLP:journals/corr/abs-1809-10361} uses coded sharding for scaling and security simultaneously.
It is based on Lagrange coded computing which provides transformation for injecting computation redundancy in unorthodox coded forms to deal with distributed failures and errors.
The key idea behind PolyShard is that instead of each node storing and processing a single uncoded shard, each node stores and computes on a coded shard of the same size that is generated by linearly mixing uncoded shards, using the well-known Lagrange polynomial.
This coding provides computation redundancy to simultaneously provide security against erroneous results from malicious nodes, which is enabled by noisy polynomial interpolation techniques (e.g., Reed-Solomon decoding).
Effectively this means that the security is as good as in traditional unsharded blockchains, which is superior to other typical sharding and delegation methods.
In PolyShard, blocks contain many transactions that have different origins.
Consequently, inter-shard communication will often be needed for verification, which is hard, and in the paper they focus on transactions that are verifiable intra-shard only.
Also, the balance summation time increases with as the chain becomes longer, which results in a total performance decrease (based on computation needs) as the chain lengths increase.

\subsection{Computing Failure Probabilities for Sharding}

Sharding raises the problem of translating the within-shard fault tolerance to a system-wide fault tolerance, and there are multiple publications that attempt to analyze the failure probabilities in sharded blockchains.
This includes a line of work by Hafid~et~al.\@~\cite{hafid2019new, hafid2020methodology, hafid2020novel, hafid2022tractable} as well as others~\cite{8418625, zamani2018rapidchain, yu2021security, rajabi2023feasibility}.
Some of these~\cite{yu2021security, rajabi2023feasibility} focus on blockchains based on proof-of-work, making them less relevant to our work.

A simple approximation (which is used in~\cite{8418625}) is to assume that a randomly chosen node has a certain probability of being Byzantine, and then use the binomial distribution to compute the probability that a shard will contain a critical number of Byzantine nodes.
Because shards do not include the same node multiple times, a hypergeometric distribution is more appropriate, as was recognized by~\cite{zamani2018rapidchain, hafid2019new, hafid2020methodology}.
(In~\cite{hafid2019new, hafid2020methodology}, Hoeffding's bound is also investigated as an approximation.)
However, this is still not an accurate model, because they analyze sharding approaches that divide nodes into disjoint sets.
The hypergeometric distribution is more accurate than the binomial distribution because choosing one node for inclusion in a sample (shard) changes the probabilities influencing the choice of the next node.
In the same way, having sampled one shard, the probabilities influencing the next shard also change.
Hafid et al.\@ addressed this in~\cite{hafid2020novel}.
In~\cite{hafid2022tractable} they reformulated the model using generator functions to also allow the probabilities to be efficiently computed.

In the case of ScaleGraph, the problem is different, because shards are not disjoint.
However, shards do not consist of random samples, in the sense that each node is not equally likely to be a member of a shard.
While the fact that shards can overlap suggests that the hypergeometric distribution could potentially be applicable, the method by which nodes map to shards suggests that it might not be.
To the best of our knowledge, the failure probability for a sharding method similar to ScaleGraph has not been modeled mathematically.
However, our results from the simulation experiment indicate that the hypergeometric distribution with an adjusted shard count is a fairly accurate approximation, and perhaps the method in~\cite{hafid2022tractable} could be applicable to ScaleGraph.

\section{Discussion and Future Work}

It is worth emphasizing again that we use the word ``shards'' to refer to subsets of nodes \emph{that often overlap}, rather than disjoint sets.
An unusual feature of ScaleGraph is that shards can overlap to an arbitrarily high degree and still offer parallelism.
The limiting factor is the number of accounts for which a single node is the leader, rather than the number of distinct subsets of nodes.

Based on the simulation experiment, we found that the choice of consensus protocol has a major impact on the shard size required to achieve a failure probability below 1 in 100,000.
These results are in line with previous work~\cite{zamani2018rapidchain, hafid2019new, hafid2020novel}.
With $f < \frac{1}{3}$ and $F=\frac{1}{4}$, even for relatively small networks, shards would need to be so large that processing one transaction per block would probably not be feasible.
On average, each node would be a member of $r$ shards, which means it would store block chains for $r$ different accounts and participate in transaction processing for $r$ different accounts.
For such large shards, such as $r > 500$, the $O(r^2)$ consensus protocol message overhead would be prohibitive.

In a permissioned setting, where participation in the network can be more controlled, targeting a smaller fraction of Byzantine nodes, such as $\frac{1}{5}$, may be realistic.
In that case, a synchronous consensus protocol allows shards to be quite small, even for large networks.
For example, with $N=16,000$ and $F=\frac{1}{5}$, a shard size $r = 81$ may be sufficient.
While targeting a larger fraction of Byzantine nodes (such as $\frac{1}{3}$) would require larger shards, the size is still reasonable and may allow adequate performance.

Assuming a permissioned setting has allowed us to side-step certain challenges that an unpermissioned environment would present, such as key generation.
We are confident that the basic principles underlying ScaleGraph could also be employed in an unpermissioned setting.

Including only a single transaction in each block allows a high degree of parallelism.
But blockchains usually contain many transactions per block for performance reasons, since, all else being equal, larger blocks enable higher throughput.
An experimental performance evaluation is crucial for determining whether (or when) the performance gain due to increased parallelism compensates for the small block size in practice.
The way validator groups are defined gives rise to many interesting corner cases.
A formal analysis proving the correctness and safety of the consensus process would be valuable in increasing confidence in the proposed solution.
In particular, the interplay between the infrastructure provided by Kademlia and the consensus protocol opens a rich set of security questions to be considered.
The security properties of Kademlia itself and the impact of Byzantine nodes, which could provide false information or drop data they are supposed to store, also deserve more attention.
Finally, for the purpose of choosing appropriate shard sizes, a more accurate mathematical model of failure probabilities would be useful.

\section{Concluding Remarks}

This paper presents a novel and scalable distributed ledger technology (DLT) method named ScaleGraph. The proposed method implements dynamic sharding using routing and logical proximity concepts from distributed hash tables (DHTs) to support frequent micro-transactions between autonomous devices. ScaleGraph combines the principles of DHTs with the principles of DLTs. It processes each transaction for consensus, involving a subset of nodes that are logically close to the sender and the receiver respectively. This allows for synchronous cross-shard verification between the sender and receiver shards directly as part of the consensus decision, which eliminates the need for hindsight cross-shard validation which is the norm in other approaches.

We analyze the critical event (transaction) ordering required for typical crypto currencies and conclude that it is sufficient to impose a per-account total order on spending events, while reception events must be ordered only relative to spending events.
This key insight is fundamental to ScaleGraph, as imposing a partial rather than total order reduces the interdependence of transactions and thus allows parallelism to be maximized.

The paper provides a high-level security analysis of the implications of sharding on availability and resistance to double spending attacks.
In general, sharding introduces a trade-off between security and performance, because larger shards make the system more resilient but less efficient.
Our simulation experiments show that an asynchronous consensus protocol would require impractically large shards.
By instead adapting a state-of-the-art synchronous consensus protocol, ScaleGraph can achieve the same resilience and failure probability with much smaller shards.

\section*{Acknowledgments}
This work has been funded by Arrowhead fPVN (grant 101111977).

\bibliographystyle{plain}
\bibliography{bibliography}

\end{document}